\documentstyle[preprint,aps,psfig]{revtex}

\begin{document}
\tighten

\def\gtwid{\mathrel{\raise.3ex\hbox{$>$\kern-.75em\lower1ex\hbox{$\sim$}}}}
\def\ltwid{\mathrel{\raise.3ex\hbox{$<$\kern-.75em\lower1ex\hbox{$\sim$}}}}
\def\bfig#1{\begin{figure}\centerline{\hbox{\psfig{file=#1}}}}
\def\efig{\end{figure}}
%%\psfigurepath{figs}
%%\psfigurepath{../thesis/figs}
%%\psfigurepath{/home/users1/pm/doug/sk/thesis/figs}
%%\psfigurepath{/home/users1/pm/doug/vms/sk/p1/figs}

\preprint{NSF-ITP-97-050\qquad gr-qc/9706011}

%%\title{Quantum Tunneling of Domain Walls in Cosmology I: Instantons}
\title{Quantum Decay of Domain Walls In Cosmology I: Instanton Approach}
\author{Shawn J. Kolitch\footnote{Electronic address:
	{\tt skolitch@pandora.physics.calpoly.edu,}
	current address: Department of Physics,
	California Polytechnic State University, San Luis Obispo, CA 93407}}
\address{Department of Physics,
	University of California,
	Santa Barbara, CA 93106-9530}
\author{Douglas M. Eardley\footnote{Electronic address:
	\tt doug@itp.ucsb.edu\hfil}}
\address{Institute for Theoretical Physics,
	University of California,
	Santa Barbara, CA 93106-4030}
\date{\today}
\maketitle
\begin{abstract}

This paper studies the decay of a large, closed domain wall in a closed
universe.  Such walls can form in the presence of a broken, discrete
symmetry.  We introduce a novel process of quantum decay for such a
wall, in which the vacuum fluctuates from one discrete state to another
throughout one half of the universe, so that the
wall decays into pure field energy.  Equivalently, the fluctuation can
be thought of as the nucleation of a second domain wall of zero size,
followed by its growth by quantum tunnelling and its collision with the
first wall, annihilating both.  The barrier factor for this quantum
tunneling is calculated by guessing and verifying a Euclidean instanton
for the two-wall system.  We also discuss the classical origin and
evolution of closed, topologically spherical domain walls in the early
universe,  through a  ``budding-off" process involving closed domain
walls larger than the Hubble radius.  This paper is the first of a
series on this subject.

\end{abstract}

\pacs{}

\section{Introduction}

Following the classic paper of Zel'dovich, Kobsarev \& Okun
\cite{ZKO}, it has been generally believed in early-universe
cosmology that domain walls are forbidden, because otherwise they would
gravitationally dominate the present-day universe, contrary to
observation.  Consequently, fundamental theories of physics are either
forbidden to exhibit spontaneously broken discrete symmetries, or are
constrained to eliminate the associated domain walls somehow, for
instance through low energy destabilization of all but one of the
discrete vacuua, or through decay of the domain walls themselves
through nucleation of string loops into walls.  Only then might an
early universe dominated by domain walls evolve into a present-day
universe consistent with current cosmological observations.

The purpose of this paper is to discuss a novel process by which domain
walls can decay, namely, the quantum decay of domain walls by global
fluctuation and quantum tunneling.  We introduce and study this process
for closed universes containing a single domain wall, but the decay
process itself is no doubt more general.  Imagine a closed universe
separated into two different discrete vacuum states by a single closed
domain wall.  If one of these vacuum states spontaneously fluctuates
into the other, the domain wall will cease to exist and energy will be
liberated.  Equivalently, the fluctuation can be thought of as the
nucleation of a second domain wall of zero size, followed by its growth
by quantum tunnelling and its collision with the first wall,
annihilating both.  Clearly this decay will be slow since it involves a
fluctuation over an entire half of the universe, and moreover it
involves curved spacetime, and therefore requires semiclassical quantum
gravity for its description.

Such a closed universe can be created, for instance, by the the
gravitational collapse of closed ({\it i.e.,} bubble-like) domain
walls.  During a cosmological phase transition that produces domain
walls by the Kibble mechanism, both open and closed domain walls will
be produced.  Closed domain walls collapse due to their surface
tension.  A collapsing closed domain wall may thermalize completely,
leaving no remnant, or it may produce a black hole by gravitational
collapse.   We especially consider in this paper the behavior of closed
domain walls that happen to be born with size $R_0$ larger than the
then-Hubble radius $R_H$.  These will originate naturally with some
probability in the Kibble mechanism.  As one might expect, such a
closed domain wall will typically produce a black hole.  However, the
spacetime singularity inside the black hole may not destroy the domain
wall;  rather, the domain wall may expand indefinitely to create a new,
inflating universe within the black hole, in a ``budding-off" process.
This process was briefly considered by Blau et al. \cite{BGG}, but has
not attracted interest since then.  Studying a simple but appropriate
model, we conclude that new-universe creation is obligatory in the
model, as long as $R_0 \gtwid R_H$ at the time of the phase transition,
and we argue that this conclusion is apt to be robust.

The newly created inflating universe is dominated by a single
closed domain wall, and therefore does not resemble our present
universe.  Since the domain wall is of finite size, however, it
is subject, as mentioned above, to decay by quantum tunneling, unlike
an infinite domain wall, which cannot decay because of an infinite
barrier against quantum tunneling.  After a long time, inflation will
therefore end by quantum tunneling of the whole domain wall, followed
by thermalization of its energy.  The resulting universe will therefore
exhibit a hot big bang, and may conceivably resemble our universe.

%% Some previous work studied the collapse of closed domain walls
%% to a black hole;
%% however this work did not cover the possibility of domain wall
%% driven inflation inside the black hole.  

A comprehensive discussion of domain walls in cosmology is given
by Vilenkin and Shellard \cite{VilShel}.
There has been a great deal of work on formation and dynamics of
closed domain walls between domains of true and false vacuua --- the
so-called ``bubble walls" --- including the possibility of budding
off of new universes within a newly formed black hole.  The present
work differs from that work in that we never appeal to a false
vacuum; all domain walls in this paper occur between regions of true
vacuum, as a result of a spontaneously broken, exact, discrete
symmetry.  There has been some very interesting work on
``topological inflation" within defects including domain walls,
cosmic strings or monopoles\cite{TopInf,LinLin}.  The present work
differs from that in that we consider quantum decay of an entire
domain wall, while that work depends on quantum fluctuations within
a domain wall, monopole, or cosmic string.  Recently,
Caldwell, Chamblin, and Gibbons \cite{andrew} studied the pair production
of black holes in the spacetime of a domain wall, using instanton
methods.  That process is related to, but distinct, from the
one we study here.

In Section II, we discuss some basic properties of domain walls and
derive the Euclidean action for walls in the thin-wall approximation.
Section III reviews the solutions describing the gravitational
fields of open (Vilenkin) and closed (Ipser-Sikivie) domain walls,
and discusses their behavior.  Section IV presents a complete
discussion of the gravitational fields and motion of domain walls
in spherical symmetry, assuming true vacuum everywhere except
for the domain wall itself.  Section V sets up a spherically
symmetric cosmological model in which a single closed domain
wall forms, and demonstrates the conditions under which this
domain wall can create a new inflating universe.  Inflation in 
this new universe is driven by the domain wall that creates it.  
Section VI contains the heart of our development and discusses
the quantum decay of the domain wall in a closed, domain wall-dominated
inflating universe, and estimates the probability per unit time
--- small but nonvanishing --- for the decay of the domain wall
by tunneling processes in semiclassical quantum gravity.

We emphasize that all the processes considered in this paper ---
budding-off, inflation, creation of cosmological fluctuations, and
quantum wall decay, are driven by domain walls.  False vacuum energy
plays no role.  This paper is the first of a series on this subject;
the second paper will study the same quantum decay process by a
different method, namely a Hamiltonian approach.

\section{The Action for Domain Walls}

Domain walls appear in matter field theories where discrete symmetries
are spontaneously broken.  The action for matter plus gravity is
of the form
\begin{equation}
	S = \int_M d^4x\sqrt{-g}\left[L_{\rm mat}
		+{R\over{16\pi G}}\right] +
		\int_{\partial M} d^3x{\sqrt{h}K_h\over{8\pi G}}.
\label{act}
\end{equation}
where the 4-volume $M$ of the system may have a 3-boundary $\partial
M${\null}, with 3-metric $h$ and 3-extrinsic curvature of trace
$K_h$.\footnote{Our conventions are generally those of MTW \cite{MTW}:
Greek spacetime indices run over $\mu,\nu,\ldots=0,1,2,3$ and the
spacetime signature is ($-$,+,+,+); Roman indices
in a hypersurface run over $a,b,\ldots=1,2,3$ or $a,b,\ldots=0,1,2$
for spacelike or timelike hypersurfaces respectively; our extrinsic
curvature $K_{ab}$ is the derivative ${1\over2}{\cal L}_{\bf n}h_{ab}$
of the hypersurface 3-metric $h_{ab}$ with respect to an outward pointing
normal vector $\bf n$.  Thus, $R>0$ and $K>0$ for a sphere embedded in
flat space.}  Little of this paper will depend on the details of the
matter, but for concreteness, the matter may be chosen as a real
scalar field $\phi$ with matter action
\begin{equation}
	L_{\rm mat} = - {1\over2}g^{\mu\nu}
		\partial_\mu\phi\partial_\nu\phi - U(\phi)
\label{actmat}
\end{equation}
where $U(\phi)$ has a discrete number of degenerate minima at which
$U=0$, {\it e.g.,} $U(\phi)$ may be a ``double well" potential.

\subsection{The Euclidean action}

The Euclidean action $I$ is obtained by analytically continuing
Eq.~(\ref{act}) to imaginary time and then reversing its sign, {\it i.e.,}
\begin{equation}
	I = -\int_M d^4x\sqrt{g}\left[L_{\rm mat}+
		{R\over{16\pi G}}\right] -
		\int_{\partial M} d^3x{\sqrt{h}K_h\over{8\pi G}}
\label{eact}
\end{equation}
where the metric is now positive definite.  We assume that $L_{\rm mat}$
has a discrete broken symmetry and therefore exhibits domain walls as
solitons in the low energy limit, without any other light degrees
of freedom.  In this paper, we treat domain walls in the ``thin-wall''
approximation \cite{SC}, where the thickness of the domain wall is
assumed much smaller than all other length scales in the problem, 
so that the wall may be taken as a two-dimensional sheet of
stress-energy in 3-space.  The low energy Euclidean action in
the thin-wall approximation is
\begin{equation}
	I_{tw} = -\int d^4x{\sqrt{g}R\over{16\pi G}} +
	\sigma\sum_i\int_{D_i} d^3x\sqrt{h} -
		\int_{\partial M} d^3x{\sqrt{h}K_h\over{8\pi G}}
\label{leact}
\end{equation}
where the sum on $i$ runs over some finite number of separate
domain walls $D_i$.  The action of each wall is simply proportional
to its 3-volume, with the surface energy density $\sigma$ a
constant, fixed by microphysics
of $L_{\rm mat}$.  In the thin-wall approximation, the spacetime
geometry is generally not smooth across each wall;  rather there
is a 3-dimensional $\delta$-function in the Riemannian curvature at
each wall \cite{WI}.  This renders the action (\ref{leact}) awkward
to work with.  Therefore, we will eliminate these $\delta$-functions
by breaking up $M$ into a union of 4-dimensional voids $M_j$, each
with smooth interior 4-geometry, meeting at domain walls $D_i$.
The Euclidean action becomes
\begin{equation}
	I_{tw} = -\sum_j \int_{M_j}' d^4x
	{\sqrt{g}R\over{16\pi G}} +
	\sum_i\int_{D_i} d^3x\sqrt{h}\left[\sigma +
	{(K_2-K_1)\over{8\pi G}}\right] -
	\int_{\partial M} d^3x{\sqrt{h}K_h\over{8\pi G}}
\label{leact2}
\end{equation}
where the prime on the integral $\int'$ means that any
$\delta$-function in $R$ at a wall does {\it not} contribute
to the 4-integral; and where $h_{ab}$ is now the 3-metric on
each domain wall $D_i$ (as well as the fixed 3-metric on the system
boundary $\partial M$), and $K_1$ and $K_2$ are the traces of the
extrinsic curvature of each domain wall on its two sides, with respect
to the normal vector that points from 1 to 2\@.  Due to the
unsmoothness of the spacetime geometry, $K_1\not=K_2$.  The instrinsic
3-metric $h_{ab}$ of each domain wall is constrained to agree with that
inherited from the 4-geometry $g_{\mu\nu}$ from the void on each of the
two sides.

The field equations become as follows.  Variation of $g_{\mu\nu}$
within each $M_j$ gives the the vacuum Einstein equations
\begin{equation}
	G_{\mu\nu} = 0 \quad\hbox{(within $M_j$)}
\label{emot1}
\end{equation}
within each void.  Variation of the domain wall metric $h_{ab}$
on each domain wall $D_i$ gives the well known Israel jump condition
\cite{WI} for the full extrinsic curvature $K^{ab}$
\begin{equation}
	K_1^{ab}-K_2^{ab} = 4\pi G\sigma h^{ab}
\label{emot2}
\end{equation}
showing that there is a jump $4\pi\sigma$ in the trace of the
extrinsic curvature at the domain wall, while there is no jump
in the traceless part of the extrinsic curvature.

\subsection{The action of a solution}

The solutions of the field equations are the extrema of the action.
After the action is varied and a solution found, simplifications
occur in the formula for value of the action at the solution.
Substituting the equations of motion (\ref{emot1},\ref{emot2}) back
into the action (\ref{leact2}) give the (weak in the Dirac sense)
formula
\begin{equation}
	I_{tw} = - \sum_i{\sigma\over2}\int_{D_i} d^3x\sqrt{h}
\label{leact3}	
\end{equation}
showing that the thin-wall action of a domain wall is always negative.

This formula was given by Caldwell, Chamblin, and Gibbons \cite{andrew}
for the particular case of a real scalar field theory (\ref{actmat}), as
follows.  The stress-energy 
tensor for the matter is obtained in the canonical way and is found to be
\begin{eqnarray}
	T_{\mu\nu}&=&-{2\over{\sqrt{-g}}}{{\delta(\sqrt{-g}L_{\rm mat})}
		\over{\delta g^{\mu\nu}}}		\nonumber\\
	&=& -\partial_\mu\phi\partial_\nu\phi + g_{\mu\nu}\left[
		{1\over2}g^{\alpha\beta}\partial_\alpha\phi
		\partial_\beta\phi + U(\phi)\right]
\label{tmunu}
\end{eqnarray}
from which one calculates, taking the trace of the Einstein equations,
\begin{equation}
	{R\over{8\pi G}}=g^{\alpha\beta}\partial_\alpha\phi
		\partial_\beta\phi + 4U(\phi),
\end{equation}
so that the Euclidean action of a solution to general relativity 
reduces to
\begin{equation}
	I = -\int\sqrt{g}d^4xU(\phi).
\label{eact2}
\end{equation}
In the thin-wall approximation, $U(\phi)=0$ except very near a wall,
where it behaves like a $\delta$-function.  In a small
neighborhood of a point on a wall, the field is approximately planar
and is only a function of the coordinate $x^3$ normal to the wall in
a Gaussian normal coordinate system.  Then the equation of motion for
the field has the first integral
\begin{equation}
	{1\over2}\left({{d\phi}\over{dx^3}}\right)^2 - U(\phi) = 0,
\end{equation}
so that Eq.~(\ref{tmunu}) gives
\begin{equation}
	T_\mu^\nu = 2U(x^3)\hbox{diag}(1,1,1,0).
\end{equation}
Furthermore, in this approximation, the stress-energy also has a 
$\delta$-function singularity at the wall \cite{IS,BGG}, so that
\begin{equation}
	T_\mu^\nu = \sigma\hbox{diag}(1,1,1,0)\delta(x^3),
\end{equation}
{\it i.e.,} $2U(x^3)=\sigma\delta(x^3)$,
where $\sigma$ is the surface energy density of the wall, as well as
its tension in the tangential directions.  Combining this result
with Eqs.~(\ref{eact}, \ref{eact2}), one again finds Eq.(\ref{leact3})
as the action of a domain wall solution to the field equations.

For convenience we will adopt the notation
\begin{equation}
	\mu\equiv 4\pi\sigma
\end{equation}
in what follows.

\section{The VIS Solutions for Wall-Dominated Universes}

A dynamical solution to Einstein's equations with a single thin
domain wall present was found by Vilenkin in \cite{AV} and by
Ipser and Sikivie in \cite{IS}.  We will henceforth refer to this
solution as the VIS solution.  One can construct this
solution from flat spacetime
\begin{equation}
	ds^2 = -dT^2 + dX^2 + dY^2 + dZ^2		\label{Mink}
\end{equation}
as follows.  The domain wall itself is the hyperboloid
\begin{equation}
	X^2 + Y^2 + Z^2 - T^2 = \left({2\over{\mu G}}\right)^2,	\label{Wall}
\end{equation}
and the ``gravitational field" of the domain wall is the interior
\begin{equation}
	X^2 + Y^2 + Z^2 - T^2 < \left({2\over{\mu G}}\right)^2.
\end{equation}
The complete solution is constructed from two such copies of flat
spacetime, glued together along their hyperboloids.  The entire
spacetime is therefore spatially closed, with topology $S^3$
in the spatial sections.  This spacetime can be interpreted as a closed
cosmology, containing a single spherical domain wall.  The universe
collapses from infinite size, halts at a minimum radius $R_{min}
={2\over{\mu G}}$ due to self-repulsion of the wall, and expands 
back out to infinite size again.

The Vilenkin solution is actually a subset of the above $S^3$ spacetime 
in a different coordinate system, and moreover it is 
conventionally given the somewhat different interpretation of an
infinite planar wall.  It is built out of the spacetime:
\begin{equation}
	ds^2 = \left({2\over{\mu G}}\right)^2 dz^2 
	+ z^2[ -dt^2 + \exp(\mu Gt) (dr^2 + r^2d\phi^2)].	\label{Alex}
\end{equation}
The complete Vilenkin solution is, similarly, two copies of this
spacetime for $z<{2\over{\mu G}}$, glued together along the domain 
wall at $z={2\over{\mu G}}$.  One can relate the two solutions as 
follows.  Rewrite Eq.\ (\ref{Mink}) 
slightly as
\begin{equation}
	ds^2 = -dU dV + dR^2 + R^2 d\phi^2		\label{Mink2}
\end{equation}
where $T=(U+V)/2$, $Z=(-U+V)/2$, $X=R\cos\phi$, and $Y=R\sin\phi$.
Then the transformation relating the solutions is:
% \begin{mathletters}
\begin{eqnarray}
      r	&=& \left(2\over{\mu G}\right){R\over V},	\\
      z	&=& \left({\mu G}\over2\right)\sqrt{R^2-UV},	\\
      t &=& \left(2\over{\mu G}\right)\ln{V \over \sqrt{R^2-UV}}.
\end{eqnarray}
% \end{mathletters}
One can verify that the Vilenkin solution given by Eq.~(\ref{Alex}) 
covers only half of the full spacetime, all of which is given by the 
Ipser and Sikivie result.  This is due to the fact that the coordinates 
in (\ref{Alex}) cover only half of the (2+1)-dimensional deSitter space
in the hypersurfaces of constant $z$.

\section{Motion of Spherical Symmetric Domain Walls}

In general, domains separated by thin walls may not consist of
entirely flat space.  In this section, we generalize the VIS
solution to include nonzero Schwarzschild masses in each
part of the spacetime, and then analyze the possible resulting
dynamics of the wall.

The junction conditions of the Einstein equations across a thin
shell of stress-energy were first worked out by Israel in \cite{WI},
who showed that the metric across the surface is continuous, whereas
the extrinsic curvature has a jump discontinuity there.  In the case
of spherically symmetric domain walls, the trajectory and 4-velocity 
of the wall are specified by
% \begin{mathletters}
\begin{eqnarray}
	x^{\alpha}(\tau)&=&(T(\tau),R(\tau),\theta,\phi)	\\
	u^{\alpha}(\tau)&=&(\dot T(\tau),\dot R(\tau), 0, 0)
\end{eqnarray}
% \end{mathletters}
and the junction condition can be written in the form
\begin{equation}
	K_{\theta\theta}^{(+)}-K_{\theta\theta}^{(-)}
		= -\mu GR^2.
\end{equation}
Taking the spacetime on each side of the wall to be Schwarzschild-deSitter, 
with line element
\begin{equation}
	ds^2 = -\left(1-{{2M}\over R}-{{\Lambda R^2}\over3}\right)dT^2
		+\left(1-{{2M}\over R}-{{\Lambda R^2}\over3}\right)^{-1}
		dR^2 + R^2d\Omega_2^2,
\end{equation}
one calculates
\begin{eqnarray}
	K_{\theta\theta}&=&R\left(1-{{2M}\over R}-{{\Lambda R^2}\over 3}
		\right)\dot T				\nonumber\\
	&=&\pm R\left(1-{{2M}\over R}-{{\Lambda R^2}\over 3}
		+\dot R^2\right)^{1\over2}		
\label{Keq}
\end{eqnarray}
where the second equality comes from normalizing the 4-velocity.  The
junction condition can then be rewritten as
\begin{eqnarray}
	\dot R^2 &=& \left[{{R(\Lambda_{+} - \Lambda_{-})}\over {6\mu}}
		+ {{(M_{+} - M_{-})}\over {\mu R^2}} \right]^2
		+ {{{(\mu GR)}^2}\over 4}		\nonumber\\
 		&&+ {{G(M_{+} + M_{-})}\over R}
		+ {{GR^2(\Lambda_{+} + \Lambda_{-})}\over 6} - 1,
\label{bigeom}
\end{eqnarray}
which is a first order ODE for $R(\tau)$, and can be regarded as an 
energy-type equation with $\mu$, $\Lambda$ and $M$ all conserved.  
The second order equation of motion for $R(\tau)$ can be found by 
differentiation.  Restricting to the true vacuum, one finds
\begin{equation}
	\dot R^2 = \left[{{(M_{+} - M_{-})}\over {\mu R^2}} \right]^2
		+ {{{(\mu GR)}^2}\over 4} + {{G(M_{+} + M_{-})}\over R} - 1,
\label{medeom}
\end{equation}
and further restricting to the case of flat spacetime, one finds
\begin{equation}
	\dot R^2 = {{{(\mu GR)}^2}\over 4} - 1,
\label{viseom}
\end{equation}
which is the VIS solution.  In what follows we will take
$\Lambda=0$, and study the behavior of domain walls in the absence
of vacuum energy, {\it i.e.,} with dynamics given by Eq.~(\ref{medeom}).

To examine Eq.~(\ref{medeom}) qualitatively, change to the dimensionless 
variables
% \begin{mathletters}
\begin{eqnarray}
	z &=& R\left[{\mu^2 G\over{2(M_{+}-M_{-})}}\right]^{1/3}	\\
	\tau' &=& {{\mu G\tau}\over 2}.
\end{eqnarray}
% \end{mathletters}
upon which the equation of motion of the wall becomes
% \begin{mathletters}
\begin{equation}
	z'^2 + V(z) = E,
\end{equation}
where
\begin{eqnarray}
	V(z) &=& -\left[{{z^6 + 2z^3(M_{+}+M_{-})/(M_{+}-M_{-}) + 1}
		\over z^4}\right],\\
	E &=& -\left[{4\over{\mu G^2 (M_{+}-M_{-})}}\right]^{2/3}.
\end{eqnarray}
% \end{mathletters}
Although we will assume $M_{+}>M_{-}$ so that $z>0$ in what follows,
this is a matter only of convention since Eq.~(\ref{medeom}) is 
symmetric with respect to $M_{+}$ and $M_{-}$.

Figure 1 shows a typical plot of $V(z)$, and one sees that there exist
several qualitatively distinct classes of solutions, depending upon the
values of the parameters $(M_{-},M_{+},\mu)$.  A wall may be born with
zero size, expand to a finite maximum radius and recollapse; it may be
born with zero size and expand indefinitely; it may collapse from
infinite size to a minimum radius and then reexpand (this is the
behavior of the VIS solution); or it may collapse from infinite size 
to zero size.  In each case, the complete spacetime consists of
two pieces of the extended Schwarzschild solution, glued together
at the wall.  

\bfig{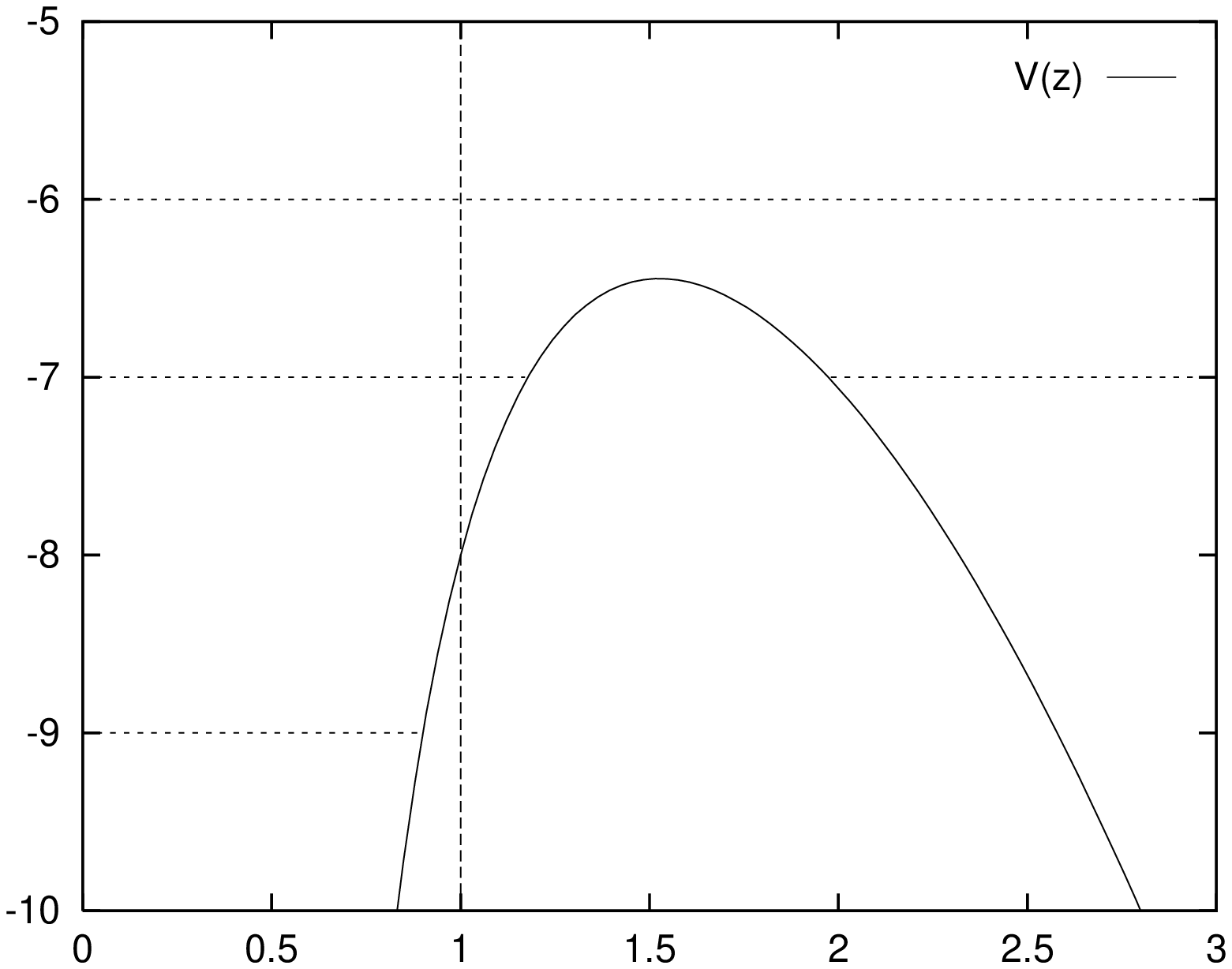,width=5.5in}
\vskip 1cm
\caption[Potential Representation of Dynamics.]{The
``potential energy'' function $V(z)$ for domain walls
moving in Schwarzschild spacetime.  When a trajectory 
crosses the vertical dashed line at $z=1$, then the 
polar angle (angle between the origin and the
trajectory) in one half of the complete spacetime
changes sign.  The horizontal lines represent the
possible qualitatively distinct classes of dynamics.}
\efig

One can construct these spacetimes as follows.  Change to coordinates
which are well-behaved at the Schwarzschild horizon, and in which
light rays travel at 45 degrees.  Then there is a correspondence
between the sign of the extrinsic curvature in Eq.~(\ref{Keq}) and
the sign of the wall trajectory's ``angular velocity'', 
${d\over{dt}}\arctan(x^0/x^1)$ in those coordinates.  For example, 
in the Kruskal-Szekeres \cite{KS} coordinates $(v, u, \theta, \phi)$, 
which are related to the standard Schwarzschild coordinates by
% \begin{mathletters}
\begin{eqnarray}
	u^2-v^2 &=& \left({R\over{2M}}-1\right)e^{R\over{2M}}	\\
	T &=&\cases{4M\hbox{arctanh}\left(v\over u\right) &$R>2M$,\cr
		    4M\hbox{arctanh}\left(u\over v\right) &$R<2M$,\cr} 
\end{eqnarray}
% \end{mathletters}
one finds 
% \begin{mathletters}
\begin{eqnarray}
	K_{\theta\theta}^{+} &=& \pm R\left(1-{{2M_{+}}\over R}
		+\dot R^2\right)^{1\over2}			\nonumber\\
		&=&+8M^2_{+}e^{-{R\over{2M_{+}}}}(u\dot v-v\dot u);\\
	K_{\theta\theta}^{-} &=& \pm R\left(1-{{2M_{-}}\over R}
		+\dot R^2\right)^{1\over2}			\nonumber\\
		&=&-8M^2_{-}e^{-{R\over{2M_{-}}}}(u\dot v-v\dot u).
\end{eqnarray}
% \end{mathletters}
In these coordinates, the polar angle in the spacetime diagram is
given by $\tan\zeta=v/u$, and its rate of change is given by
\begin{equation}
	\dot\zeta={{u\dot v-v\dot u}\over{u^2+v^2}}.
\end{equation}
Hence we see that when $K_{\theta\theta}^{+}$ is positive in a given
region of the spacetime, the wall trajectory follows a path such that
$\zeta$ increases along the trajectory, and vice versa.  Similarly
one sees that when $K_{\theta\theta}^{-}$ is positive, $\zeta$
{\it decreases} along the wall trajectory, and vice versa.  The sign
change is due to the fact that the vector normal to the 4-velocity 
changes its orientation from one side of the wall to the other. 

The last step is to determine the signs of the quantities 
$K_{\theta\theta}^{+}$ and $K_{\theta\theta}^{-}$ as a function
of $z$ in Fig. 1.  One finds that
% \begin{mathletters}
\begin{equation}
	K_{\theta\theta}^{+}\cases{>0,	&$z>1$\cr
				   <0,	&$z<1$\cr}
\end{equation}
whereas
\begin{equation}
	K_{\theta\theta}^{-} <0,  \hskip 0.5cm z>0.
\end{equation}
% \end{mathletters}
Finally one knows everything necessary to construct the complete
spacetimes from Fig. 1.  In each part of the spacetime, one only has
to draw the wall trajectory such that it has the correct starting
and ending points, and also so that the angle $\zeta$ increases or
decreases in each part according to the signs of the extrinsic 
curvature as stated above.

Figure 2 shows the Penrose (conformal) spacetime diagrams for each of 
the possible cases.  In each case, the complete spacetime consists
of the wall trajectory, those points in the right-hand copy of 
Schwarzschild to its right, and those points in the left-hand copy of
Schwarzschild to its left.  Of particular interest is case (iv), where 
a wall is born with zero size and grows to infinite size.  An observer
in the asymptotically flat region IV of the spacetime would see the 
wall enter the Schwarzschild horizon; however, the wall subsequently 
avoids the singularity within the horizon and expands indefinitely, 
creating a new, inflating universe inside the black hole.  Clearly, 
this model suffers from the requirement that an initial singularity 
exist from which the domain wall emerges.  We therefore proceed 
to explicitly construct a model which circumvents this necessity, 
while still producing a new wall-dominated universe. 

\bfig{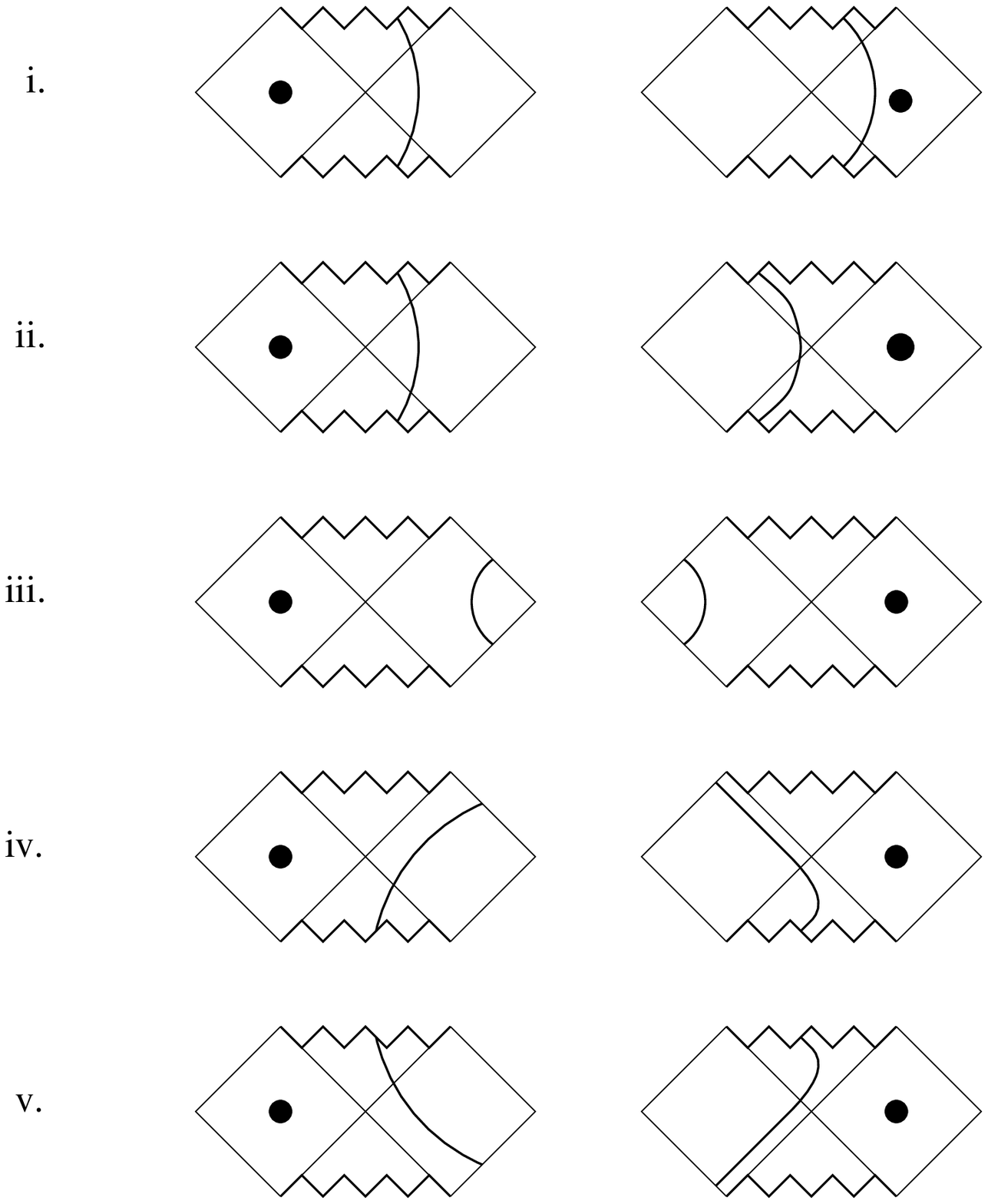,width=5.2in}
\vskip 0.1cm
\caption[Penrose Diagrams.]{Spacetime diagrams for
each of the possible qualitatively distinct classes
of dynamics, for domain walls moving in Schwarzschild
spacetime.  In each case, the complete spacetime
consists of the wall, those points in the right-hand
copy of Schwarzschild to its right, and those points
in the left-hand copy of Schwarzschild to its left.}
\efig

\section{A model of newly formed bubbles of domain wall}

\def\Eq#1{Eq.\ (\ref{#1})}
\def\Eqs#1{Eqs.\ (\ref{#1})}
\def\DM{\Delta M}
\let\s=\mu
\def\e{\epsilon}
\def\1#1{{1\over#1}}
\def\sm{\sqrt{1+8m^2}}
%%\singlespace

We now set up and study a toy model of the behavior of topologically
closed and spherical domain walls in cosmology.  For convenience we
work in units where $G=1$, in this Section only.  Our cosmological
model is spherically symmetrical, closed, and bounded.
The matter content of the model is pressureless dust, arranged in
concentric spherical shells.   An important feature of our model is
that would be fated to recollapse in a finite time, if it were not for
the newly born closed domain wall.  This feature is important because we
want to avoid giving the newborn domain wall a spurious kick,
{\it i.e.,} a positive total energy.  In applications we may think of the
model as being only moderately larger than the Planck radius, and
with a total lifetime only moderately larger than the Planck time
--- if it were not for domain wall effects.

(As a limiting case we can study a universe in which each shell starts
with marginal (``$k=0$") binding energy, so that the whole universe just
barely expands to infinity as $t\rightarrow\infty$.)

In other words, we want our model to be a {\it purely curvature-free
pertubation} of a completely homogeneous pressureless $k=+1$ model,
because phase transitions cannot produce correlated pertubations
outside the light cone, and therefore, as is well known, cannot produce
curvature perturbations.

We concentrate on a single dust shell of finite total proper mass
$\DM$, idealized as being infinitely thin. \footnote{An infinitely
thin ``dust" shell of finite mass is an acceptable idealization within
general relativity;  however, subtleties do lurk:  A close analysis
shows that the shell must possess radial stresses in order to stay
infinitely thin --- this is for instance clear from from the fact that
a dust particle just inside the shell must experience a different
acceleration than one just outside.} Spacetime is vacuum for a small
region (see below) around our shell, to avoid complications due to
collisions of shells;  see below.  By Birkhoff's theorem, the spacetime
just inside the shell is Schwarzschild of some mass $M_1$, and that
outside is Schwarzschild of some mass $M_2$.  Our shell moves according
to the first-order equation of motion
\begin{equation}
\dot R^2 = {\DM^2\over 4R^2}-1 + {M_2+M_1\over R} +{(M_2-M_1)^2\over\DM^2}
\end{equation}
where $R(\tau)$ is the curvature radius of our shell as a function of
its proper time $\tau$, and $\dot\empty = d/d\tau$.  In order that the
shell be gravitationally bound, with some relative binding energy $\e$,
we take
\begin{equation}
 M_2-M_1 = \e\DM \quad (0<\e\le1).
\end{equation}
Then the equation of motion becomes
\begin{equation}
\dot R^2 = {\DM^2\over 4R^2} + {M_2+M_1\over R} - 1 + \e^2.
\end{equation}
If $0<\e<1$ our shell would recollapse; in the limit
that $\e=1$ our shell would just make it to infinity --- if left
undisturbed.  However, it is instead going to turn into a domain wall.

To model the birth of a domain wall, our shell is assumed to suddenly
become a domain wall as it reaches some radius $R_0$:  a tangental
stress equal in magnitude to the surface mass density suddenly appears.
However, the mass density (and total mass $\DM$) must remain meanwhile
remain constant, according to conservation laws.  Thereafter, our
shell-cum-domain-wall moves according to a different first-order
equation of motion,
\begin{equation}
 \dot R^2 = \14\s^2R^2-1+{(M_2+M_1)\over R}+{(M_2-M_1)^2\over\s^2R^4} 
\end{equation}
To match $R$ and $\dot R$ at $R_0$ (these must be continuous) we have
\begin{equation}
	\s = {\DM\over R_0^2}
\end{equation}

Thus the first-order equation of motion of $R$ for the spherical domain
wall becomes
\begin{equation}
 \dot R^2 = {\s^2R^2\over4} -1 +{M_2+M_1\over R}+{\e^2R_0^4\over R^4}
\end{equation}
The main question is simple:  How does the solution of this equation 
$R(\tau)$ behave?  Does it go to $\infty$ as $\tau\rightarrow\infty$,
or in contrast does it fall back to $R=0$ at some finite time $\tau$?
The former behavior means that the shell has formed a new, inflating,
semi-closed universe inside the black hole $M_2$, while the latter
behavior means that the shell has crunched into the singularity inside
the black hole $M_2$.  Note that in both cases, the domain wall creates
a new black hole of mass $M_2$ by gravitational collapse, as seen
by external observers in the original universe.

Our model therefore has three dimensionful paramaters, $\mu$, $M_1$,
and $R_0$, and one further dimensionless parameter, $\e$.
(By the above, $M_2 = M_1 + \s\e R^2$.)  Physically, $\mu$ is fixed
by microphysics, $R_0$ is set through the Kibble mechanism by random
variations during the phase transition and is likely to be roughly the
Hubble radius then, and $M_1$ is the mass of matter that happens to lie
inside the bubble of domain wall when it forms.  The relative binding
energy $\e$ is set by the overall dynamics of the initial universe,
with some further adjustment by the Kibble mechanism. 

Since three of the parameters are dimensionful,
it is convenient to render them dimensionless by forming ratios with
the quantity $M_2+M_1$, reducing them to two dimensionless parameters,
$m$ and $r_0$, with $0\le m \le 1$ and $0<r_0<\infty$:
\begin{eqnarray}
     m	&=& {M_2-M_1\over M_2+M_1} = {\e\s R_0^2\over \s\e R_0^2 + 2M_1},\\
    r_0	&=& {R_0\over M_2+M_1} =      {R_0\over \s\e R_0^2 + 2M_1};\\
\noalign{\noindent and a dimensionless radius $r(t)$ as a function of
a dimensionless proper time $t$,}
     r &=& {R\over M_2+M_1}	=  {R\over \s\e R_0^2 + 2M_1},\\
     t &=& {\tau\over M_2+M_1} = {\tau\over \s\e R_0^2 + 2M_1}.
\end{eqnarray}

In terms of the dimensionless quantities, the equation of motion is
\begin{equation}
	{r'}^2 + V(r) = 0
\end{equation}
where the effective potential is
\begin{equation}
 V(r) \equiv -{m^2\over4\e^2r_0^4}r^2 + 1 - {1\over r} -
			{\e^2r_0^4\over r^4}.
\end{equation}
The effective potential $V(r)$ diverges to $-\infty$ as either
r$\rightarrow0$ or $r\rightarrow\infty$.  Therefore $V(r)$ has a
maximum at some point $r_x$ which will be studied below;  it will turn
out that $V(r)$ has only one local maximum for $0<r<\infty$.  If the
system point gets to sufficiently large $r$, a new, semi-closed, inflating
universe will be born.  The system point starts at $r=r_0$, at the moment
when the domain wall is born.  The key question is whether it can get past
the maximum at $r_x$.  It will get to large $r$ if either of two
conditions is satisfied:  1) $r_0 \ge r_x$; or 2) $V(r_x)<0$.

Let us find the maximum of $V(r)$.  The gradient of $V$ is
\begin{equation}
	dV/dr = -{m^2\over2\e^2r_0^4}r +1/r^2 + {4\e^2r_0^4\over r^5}
\end{equation}
and the only positive root of the equation $dV/dr=0$ is found to be
\begin{equation}
	r_x = {r_0^{4/3}\e^{2/3}\over m^{2/3}}\left(1+\sm\right)^{1/3}.
\end{equation}
The condition 1) above that $r_0\ge r_x$ is then
\begin{equation}
	r_0 \le {m^2\over\e^2(1+\sm)}		\label{cond1}
\end{equation}
The maximum value of the potential is
\begin{equation}
	V(r_x) = 1-{3m^{2/3}\over2\e^{2/3}r_0^{4/3}}{1+2m^2+\sm\over
		\left(1+\sm\right)^{4/3}},
\end{equation}
and condition 2) that this maximum be negative is
\begin{equation}
  r_0 < \left({3\over2}\right)^{3/4} {m^{1/2}\over\e^{1/2}}
	{\left(1+2m^2+\sm\right)^{3/4}\over1+\sm}	\label{cond2}
\end{equation}
Only one of Eqs.\ (\ref{cond1}, \ref{cond2}) need be satisfied for the
system to get to large $r$.  Comparing these equations one sees that
\Eq{cond2} dominates ({\it i.e.,} is less restrictive than) \Eq{cond1} 
as long as
\begin{equation}
  \e > \left({2\over3}\right)^{1/2}{m\over(1+2m^2+\sm)^{1/2}} \label{dom}
\end{equation}
while on the other hand \Eq{cond1} dominates if this inequality is
reversed.  In particular, \Eq{dom} holds for all permissible $m$ if
$\e>1/3$.  So let us concentrate on the case $\e>1/3$ (shell not
greatly bound) and study \Eq{cond2}.

Returning to dimensional variables, the condition \Eq{cond2} becomes
\begin{equation}
  \e\s^2R_0^2 + 2\s M_1 >
	\left({2\over3}\right)^{3/2}{\left(\sm\right)^2\over
	\left(1+2m^2+\sm\right)^{3/2}}.		\label{inflate2}
\end{equation}
This equation is the condition that the newly formed bubble will form
an inflating, semi-closed universe.  The RHS of \Eq{inflate2} is of
order unity and nearly constant for $0\le m\le1$, decreasing from 0.77
at $m=0$ to 0.59 at $m=1$.  In particular, this
equation shows that such a universe will {\it always} form if the
initial bubble radius $R_0$ is large enough,
\begin{equation}
	\s R_0 \gtwid 0.8/\e^{1/2}.
\end{equation}
Furthermore, \Eq{inflate2} can be shown to imply that the bubble is
outside the Hubble radius when it forms, in the sense that $2M_2/R<1$.

Equation (\ref{inflate2}) is also sufficient to form a new inflating
universe even if $\e$ is small and \Eq{dom} is not satisfied, but it
is not necessary.  In that case it is easier to form a new inflating
universe than implied by \Eq{inflate2}.

\section{Instantons and quantum decay}

We now wish to estimate the probability per unit time that the 
wall-dominated universe we have described in the previous section 
will decay due to quantum effects.  For simplicity, we will 
consider the case of the VIS solution; however, in the absence 
of false vacuum energy, one would expect the effect of a nonvanishing 
Schwarzschild mass to be small.  We proceed by constructing a 
Euclidean instanton which can be interpreted as interpolating 
between a time slice of the VIS solution and a time slice 
of a two-wall spacetime in which a second domain wall has 
nucleated at zero size, with the two walls tunnelling toward 
each other until they collide and annihilate.

We begin by calculating the action of the Euclidean VIS solution.  Recall 
that the radius of the wall as a function of the Minkowski time in flat 
coordinates is given by Eq.~(\ref{Wall}):
\begin{equation}
	X^2 + Y^2 + Z^2 - T^2 = \left({2\over{\mu G}}\right)^2,	
\end{equation}
and that the complete spacetime consists of the interiors of two such
hyperboloids, identified at their boundaries (the wall).
The Euclidean VIS solution therefore consists of 2 balls of flat
4-space, with the wall as their common (identified) boundary.  Then
by Eq.~(\ref{leact3}), the action of this solution is 
\begin{eqnarray}
	I_{VIS} &=& -{\mu\over{8\pi}}\int\sqrt{h}d^3x	\nonumber\\
	&=& -{\mu\over{8\pi}}(2\pi^2 a^3)	\nonumber\\
	&=& -{{2\pi}\over{\mu^2 G^3}}.
\end{eqnarray}

Having constructed the Euclidean VIS thin-wall solution, we can now
construct a more general class of thin-wall Euclidean spacetimes, some
members of which are solutions to the field equations, and therefore
candidates for instantons that mediate interesting processes such as
quantum tunneling.  We construct these spacetimes out of some even
number $2n$ ($n=1,2,3\ldots$) of pieces; the Euclidean VIS solution,
composed of two pieces, will be the case $n=1$. Each piece is a
lens-shaped region of flat 4-space bounded by two 3-spherical segments.
 The two 3-spherical segments join on a complete 2-sphere (``the edge
of the lens").  We form the complete space by identifying pairwise the
3-spherical segments in round-robin fashion;  all the 2-spherical edges
are identified to a single 2-sphere.  We demand that space be locally
flat at this 2-sphere (no conical singularity --- angle  of $2\pi$
around any circle).   In the whole space, the walls are precisely the
$2n$ identified 3-spherical segments.  It will now be shown that for a
given value of $n$, a configuration which extremizes the  Euclidean
action consists of $2n$ identical lenses, with the two  3-spherical
segments bounding each lens meeting at an angle $2\pi/2n$. Furthermore,
the radius of curvature of each segment will be equal to $2/\mu G$, the
radius of the VIS solution.

For a flat, compact spacetime, the action in the form of
Eq.~(\ref{leact2}) becomes
\begin{equation}
	I_{tw} = -\sum_i\int_{D_i} d^3x\sqrt{h}\left[{-(K_1+K_2)
	+ 2\mu\over{8\pi G}}\right],
\end{equation}
where the sum is over the $2n$ 3-spherical lens boundaries as
described above.  Each 3-spherical segment can be described as 
a segment, or cap, of $S^3$ of some fixed radius $a_i$ and maximum
polar angle $\theta_i$:
\begin{equation}
	ds^2 = a_i^2\left(d\theta^2 + \sin^2\theta d\Omega^2\right),
\end{equation}
where $0\le\theta\le\theta_i$.

The extrinsic curvature and its trace at the surface of a 
3-sphere of radius $a$ are given by $K_{11}=K_{22}=K_{33}=\pm 1/a$ 
and $K = \pm 3/a$, where the sign depends on whether the sphere 
is chosen to have positive or negative curvature on each side.  
The case of a positive-energy domain wall corresponds to choosing
negative curvature on each side of the wall, {\it i.e.,} an
observer on either side of the spherical wall is enclosed by it.  
Using $K_1+K_2=-6/a$ in the above action, we arrive at
\begin{equation}
	I_{tw} = {1\over{4\pi}}\sum_i\int_{D_i} d^3x\sqrt{h}\left[{-3
		\over{Ga_i}} + \mu\right].
\end{equation}
In order that the geometry be smooth,
we demand that the area of the 2-sphere which joins any pair of 
3-sperical segments be the same for all pairs of segments, hence 
\begin{equation}
	A_2 = \int\sqrt{g_{\chi\chi}g_{\phi\phi}}d\chi d\phi
		= 4\pi a_i^2\sin^2\theta_i = {\rm const}
		\equiv 4\pi a^2,
\label{radconstr}
\end{equation}
where $a$ is a constant for all lenses in a configuration.  Likewise
we demand smoothness
(no conical singularity) at the pole of our configuration; hence
\begin{equation}
	\sum_{i=1}^{2n}\theta_i = \pi.
\label{angconstr}
\end{equation}
These constraints can be used to eliminate the $2n$ radii $a_i$
in favor of the single parameter $a$, and to eliminate one of the
polar angles in favor of the remaining $2n-1$.  The physical parameter
space is now $(0<\theta_i<\pi,\ a>0)$, where $i=1,\dots,2n-1$.

The action is then
\begin{eqnarray}
	I_n(\theta_1,\dots,&&\theta_{2n-1},a) =		\nonumber\\ 
		&&{1\over2}\sum_{i=1}^{2n-1}\left(
		\theta_i-{{\sin2\theta_i}\over2}\right)
		\left({{\mu a^3}\over{\sin^3\theta_i}}
		-{{3a^2}\over{G\sin^2\theta_i}}\right)	\\
		&&+{1\over2}\left(\pi-\sum\theta_i+
		{{\sin2\sum\theta_i}\over2}\right)
		\left({{\mu a^3}\over{\sin^3\sum\theta_i}}
		-{{3a^2}\over{G\sin^2\sum\theta_i}}\right).\nonumber
\end{eqnarray}
To extremize the action over our configurations, we set
% \begin{mathletters}
\label{exteq}
\begin{eqnarray}
	0={{\partial I_n}\over{\partial\theta_i}} &=& 
	\sin^2\theta_i\left[\mu\left({a\over{\sin\theta_i}}\right)^3
	-{3\over G}\left({a\over{\sin\theta_i}}\right)^2\right]	
						\nonumber	\\
	&+&{{3\cos\theta_i}\over{2\sin\theta_i}}\left(\theta_i
	-{{\sin 2\theta_i}\over 2}\right)\left[{2\over G}
	\left({a\over{\sin\theta_i}}\right)^2-\mu
	\left({a\over{\sin\theta_i}}\right)^3\right]		
						\nonumber	\\
	&-&\sin^2\Sigma\theta_i\left[\mu
	\left({a\over{\sin\sum\theta_i}}
	\right)^3-{3\over G}\left({a\over{\sin\sum
	\theta_i}}\right)^2\right]				
						\\
	&+&{{3\cos\sum\theta_i}\over{2\sin
	\sum\theta_i}}\left(\pi-\sum_{i=1}^{2n-1}
	\theta_i-{{\sin 2\sum\theta_i}\over 2}\right)		
						\nonumber\\
	&\times&\left[{2\over G}\left({a\over{\sin\sum\theta_i}}\right)^2
	-\mu\left({a\over{\sum\sin\theta_i}}\right)^3\right],
						\nonumber
\end{eqnarray}
and
\begin{eqnarray}
	0={{\partial I_n}\over{\partial a}} &=& 
	{3\over2}\sum_{i=1}^{2n-1}\left({{\theta_i-
	{1\over2}\sin 2\theta_i}\over{\sin\theta_i}}\right)
	\left[\mu\left({a\over{\sin\theta_i}}\right)^2
	-{2\over G}\left({a\over{\sin\theta_i}}\right)\right]	
						\\
	&+&{3\over2}\left({{\pi-\sum\theta_i+
	{1\over2}\sin 2\sum\theta_i}\over{\sin\sum\theta_i}}\right)
	\left[\mu\left({a\over{\sin\sum\theta_i}}\right)^2
	-{2\over G}\left({a\over{\sin\sum\theta_i}}\right)\right],
						\nonumber
\end{eqnarray}
% \end{mathletters}
which has the obvious, symmetrical solution
% \begin{mathletters}
\begin{eqnarray}
\label{symmsoln}
	&&\theta_1=\dots=\theta_{2n-1}={\pi\over {2n}},		\\
	&&{a\over {\sin\theta_i}}={2\over{\mu G}}.
\end{eqnarray}
% \end{mathletters}
We believe that there are no other allowed solutions, but have
not proved this.  Although we have extremized the action only
within our set of configurations, in fact the extremizing
spacetimes are solutions of the full Euclidean field equations.
Next we determine whether these extrema are minima, maxima
or saddlepoints of the action.

To determine the character of an extremum of a function of several
variables, one constructs the Hessian matrix, or matrix of second
partial derivatives.  Then the extremum is a local minimum if and
only if every eigenvalue of that matrix is nonnegative, whereas each
negative eigenvalue represents a direction in which the function
decreases, so that if such eigenvalues exist then the extremum is
a saddlepoint \cite{webb}.  In the present case one finds
% \begin{mathletters}
\begin{eqnarray}
	H_{\theta_i\theta_j}&=&A\times
		\pmatrix{-2&-1&-1&\ldots&-1\cr
			 -1&-2&-1&\ldots&-1\cr
			 \vdots&\vdots&\vdots&\vdots&\vdots\cr
			 -1&-1&\ldots&-1&-2},			\\
	H_{aa}&=&B,						\\
	H_{\theta_i a}&=&0,
\end{eqnarray}
% \end{mathletters}
where $A$ and $B$ are positive constants for a given value of $n$,
defined by
% \begin{mathletters}
\begin{eqnarray}
	A &\equiv& {2\over{\mu^2 G^3}}\left[2\sin{\pi\over n}-
		3\cot^2{\pi\over{2n}}\left({\pi\over n}-\sin
		{\pi\over n}\right)\right],			\\
	B &\equiv& {3\over G}\left[{{\pi-n\sin{\pi\over n}}\over
		{\sin^2{\pi\over{2n}}}}\right].
\end{eqnarray}
% \end{mathletters}
The angular part of the Hessian has eigenvalues $(-2n,-1,\dots,-1)$,
so that this extremum is clearly a saddlepoint of the Euclidean
action.  However, since no other extrema appear to exist in the
physically allowed parameter space, this symmetric solution can still
be interpreted as an instanton which
mediates quantum tunneling.\footnote{In general, it is not possible
to determine absolutely that stable solutions to the nonlinear
Eqs.~(\ref{exteq}) do not exist in the physical parameter space
$(0<\theta_i<\pi,\ a>0)$, {\it i.e.,} that a local minimum of the
action does not exist and make the dominant contribution to the
Euclidean path integral for the tunneling amplitude.
However, we have searched for such solutions numerically, and we
do not believe that they exist.}

The total Euclidean action of this solution is 
\begin{equation}
	I_n = -{{2\pi}\over{\mu^2 G^3}}
		\left(1-{n\over\pi}\sin{\pi\over n}\right).
\end{equation}
Since this action is an increasing function of $n$, the VIS solution
$(n=1)$ is the thin-wall solution of least action, and the next lowest 
lying solution will be $n=2$, which we therefore choose as our
candidate instanton.  It consists of 4 lenses of flat 4-space, bounded
by segments of 3-spheres which we identify pairwise and each pair of
which meets at the angle $\pi/2$.  One can think of this instanton
as mediating the creation from nothing of two domain walls;
from Eq.~(\ref{radconstr}), each wall has radius of curvature 
$a_2=a_1\sin(\pi/4)$, where $a_1=2/(\mu G)$ is the VIS radius.
Hence the two walls are created by the instanton with radius
$R=\sqrt{2}/(\mu G)$.

One can choose a ``final'' 
slice of zero extrinsic curvature through this 4-geometry 
such that at an instant of Euclidean time, a single wall separates 
two domains which contain the same phase of the scalar field;  
the slice passes through the centers of two of the lenses as shown
in Figure 3.  Let us refer to these two lenses as the primary 
lenses.  Then the slices evolve backwards in Euclidean time
towards the VIS solution as follows:  successive slices intersect 
the primary lenses in a sequence of 2-spheres of decreasing radii, 
with each slice passing through one of the intermediate lenses in 
such a way that the 2-spheres shrink faster in one of the primary 
lenses than in the other.  This process continues until the smaller 
of the 2-spherical slices has reached zero radius, as shown in
Figure 4; this is the point at which the second wall appears.
We require that the initial slice be isomorphic to one of zero 
extrinsic curvature through the VIS space; however, our instanton
contains no such slice.  

\bfig{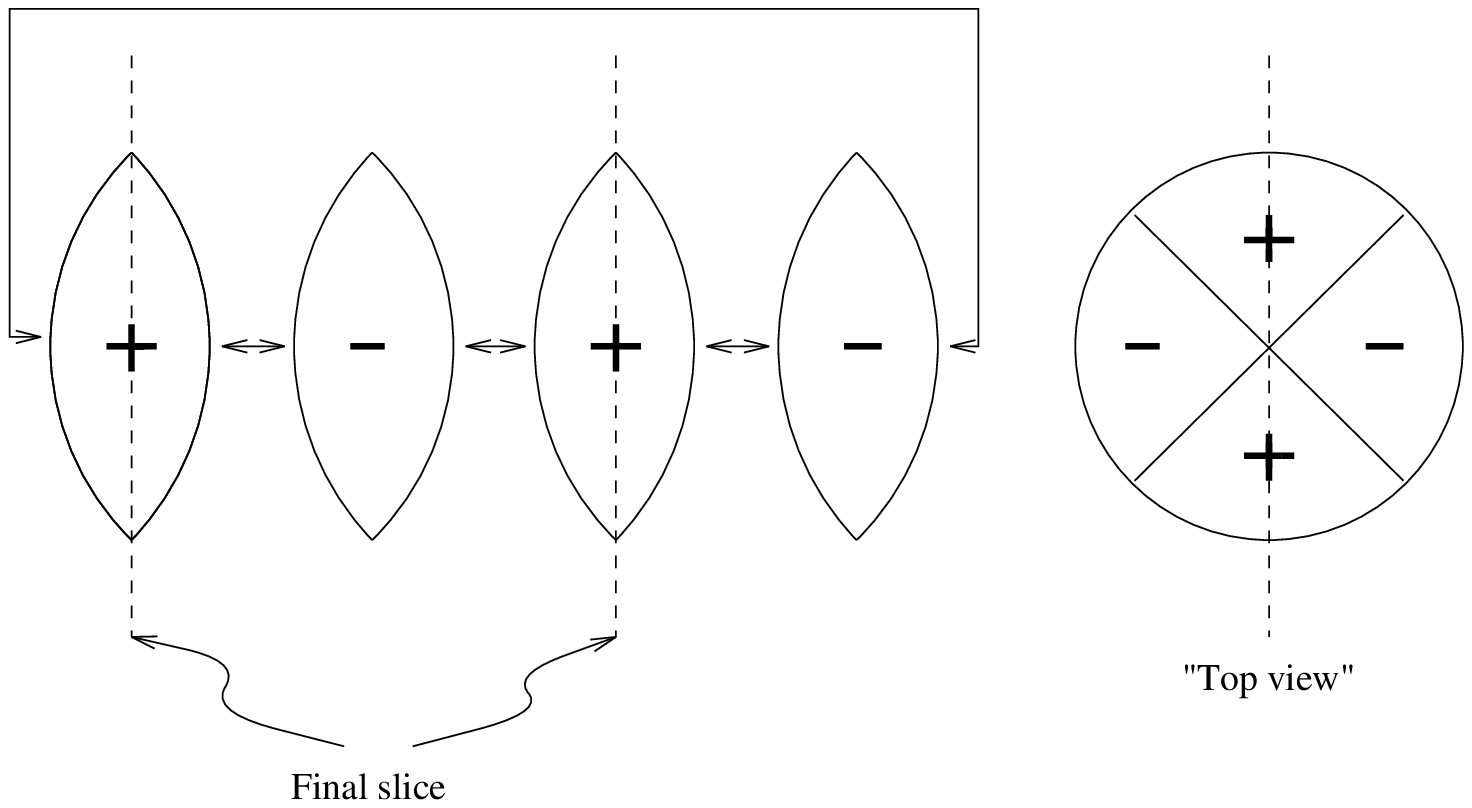,width=5.5in}
\vskip 1cm
\caption[The Final Hypersurface.]{A hypersurface of zero 
extrinsic curvature passing through the $n=2$ symmetric
instanton.  The surface contains two regions of space
where the phase of the scalar field is the same.}
\efig

\bfig{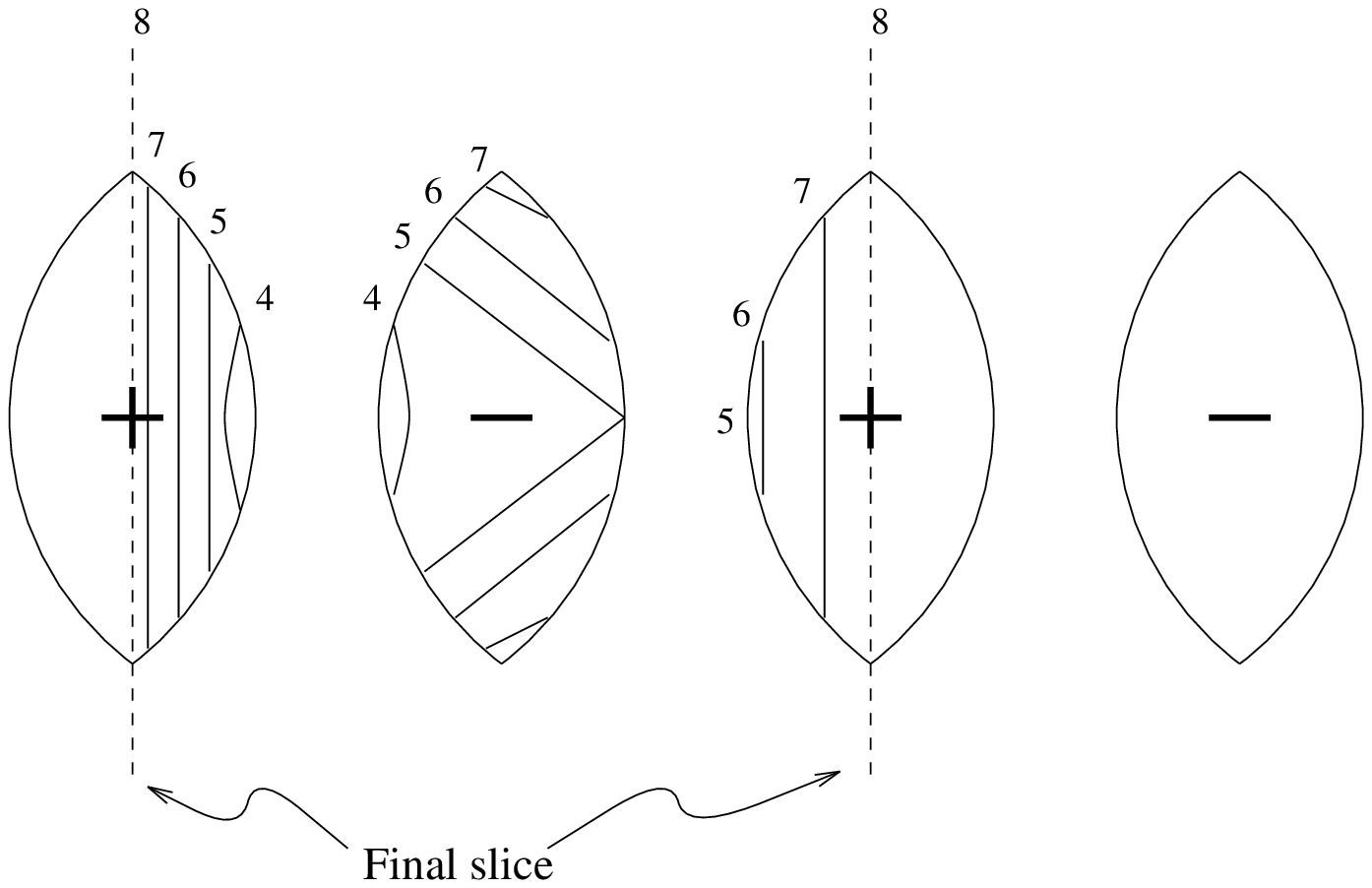,width=5.5in}
\vskip 1cm
\caption[2nd Half of the Evolution.]{The final stages of the
evolution of the hypersurface which interpolates between
the final and initial configurations.  Successive instants
of Euclidean time are numbered; slice 5 represents the instant
at which the second wall is created with zero radius.}
\efig

Although the meaning of this is unclear, 
similar pathologies have been encountered in Euclidean quantum 
gravity before.  One way of getting around the difficulty is
simply to adopt the no-boundary proposal, {\it i.e.,} to calculate
the barrier factors for tunneling from nothing of the initial
and final field configurations, and then take their difference as 
the action of the interpolation.  This was the approach taken, for
example, in \cite{MJP1,MJP2,andrew}.  Another approach has been 
proposed by Farhi, Guth and Guven \cite{FGG}, who encountered this
difficulty when calculating the amplitude for a false-vacuum 
bubble to tunnel through a classically forbidden region, so that
it could expand perpetually.  They suggested the following rule:
that the Euclidean manifold exhibits a 2-sheeted structure, with
the interpolating slices moving for part of their evolution on the
second sheet, which would then contain
a suitable initial slice.  (One must then find a single slice which 
can be matched between the two sheets.)  In their work, they called 
such a multi-sheeted Euclidean manifold a ``pseudomanifold'',
\footnote{Not to be confused with a pseudomanifold in topology.}
and they defined the covering number of any point in the manifold
as the number of times the point is crossed by the evolving 
hypersurface in the future direction, minus the number of times 
it is crossed in the past direction.  They then found that the 
action weighted by the covering number yields the correct Euclidean 
equations of motion.  

In the present case the second sheet, on which the slices evolve 
before the new domain wall has appeared, is simply the Euclidean VIS
manifold.  The evolution then proceeds as shown in Figure 5.  
The action of the interpolation is calculated by following
the hypersurface through its complete evolution, and summing the
wall area (and thus the action) ``swept out'' by it, weighted by 
the covering number as described above.  For the first part of the 
evolution (slices 1--4 in Fig.~5), the action is
\begin{equation}
	{1\over2}I_{VIS} - \Delta I,
\end{equation}
where $I_{VIS}$ is the Euclidean action of the VIS solution,
and $\Delta I$ is the action of the area of the 3-hemisphere 
{\it not} swept out by this part of the evolution.  Now the 
slice begins evolving on that part of the manifold containing 
the instanton, and the action of this part of the interpolation
is
\begin{equation}
	-\left({1\over2}I_2 - \Delta I\right),
\end{equation}
where $I_2$ is the Euclidean action of the instanton, and the
minus sign is present because the slice evolves through these
points with the opposite orientation.  Summing the two
contributions, we see that the total Euclidean action of the 
interpolation is
\begin{equation}
	I = {1\over2}(I_{VIS}-I_2).
\end{equation}

\bfig{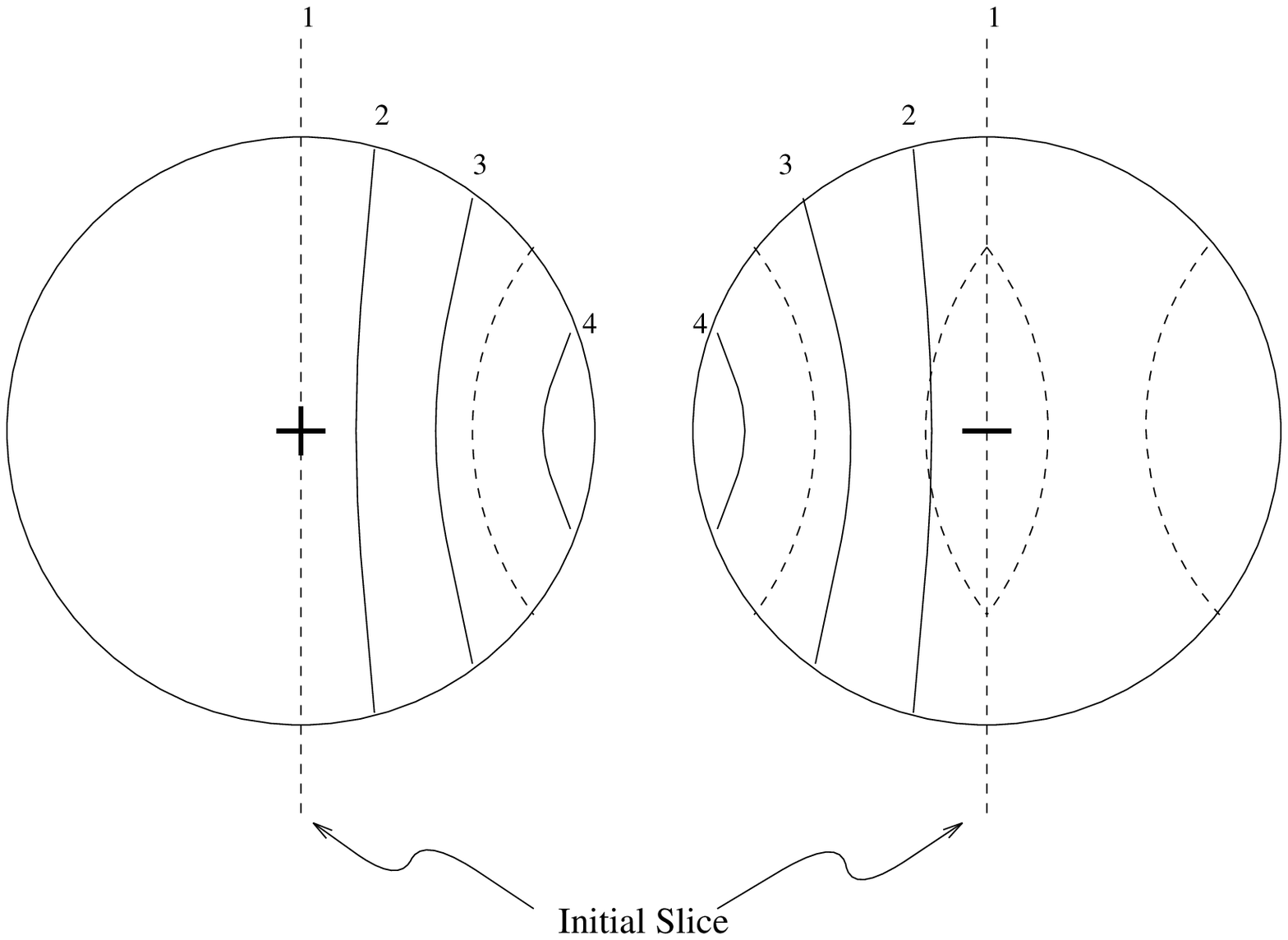,width=5.5in}
\vskip 1cm
\caption[1st Half of the Evolution.]{The initial stages of the
evolution of the interpolating hypersurface, which take place
on the VIS sheet of the manifold.  The second sheet, consisting
of the $n=2$ instanton, is indicated by the curved, dashed lines.}
\efig

A physical interpretation
of this process is that in one half of the VIS space, a second wall 
is nucleated quantum mechanically and grows in imaginary time
while the first wall shrinks; when the walls reach the same
place and same size, they annihilate in a burst of energy.  In other
words, one imagines the scalar field in one half of the space
fluctuating, in its entirety, into the opposite vacuum state.
At the same time, the geometry fluctuates so that the domain
wall radius is reduced by a factor of $\sqrt{2}$ at the instant
of its decay.

The Euclidean action of the instanton is
\begin{eqnarray}
	I_2 &=& {{-2\pi}\over{\mu^2 G^3}}
		\left(1-{2\over\pi}\sin{\pi\over 2}\right)	
							\nonumber\\
	&=& {{4-2\pi}\over{\mu^2 G^3}}.
\end{eqnarray}		
The action of the interpolation is therefore
\begin{equation}
	I={1\over2}(I_{VIS}-I_2)=-{2\over{\mu^2 G^3}}.
\end{equation}
The amplitude for the decay process is then given in the semiclassical 
limit by \cite{CDL}
\begin{equation}
	{\Gamma\over V}=A\exp[-I],
\end{equation}
where the prefactor $A$ can by calculated by considering perturbations
about the instanton \cite{SC}.  Ignoring the prefactor, we thus find 
that the probability for decay will contain the barrier factor
\begin{equation}
	P \sim \exp\left[-{2\over{\mu^2 G^3}}\right].
\end{equation}

\section{Conclusions}

The study of domain walls in the early universe is motivated
both by their inflationary nature, and by the naturalness with 
which they arise from fundamental field theories.  The spacetimes
associated with domain walls are in many ways analagous to deSitter
spacetime, with the vacuum energy confined to the two-dimensional
sheets which are the walls.  However, the presence of domain walls
does not require a global source of vacuum energy; only a broken
discrete symmetry need exist to lead to the formation of the walls.
Domain walls may arise more naturally in microphysics than the false
vacuum which commonly drives models of inflation.

As in the case of false vacuum energy, domain walls tend to
dominate the classical dynamics of the universe due to their 
gravitational properties.  However, the resulting cosmological 
models exhibit large-scale anisotropies in the CMBR, inconsistent 
with observations, and although the instability of the false vacuum 
is well-known \cite{CDL}, there was until now no corresponding result 
in the case of pure domain walls.  For this reason, it has long been 
believed that domain walls should be forbidden in the early universe.  
The primary result of this paper has been to demonstrate that closed
domain walls are in fact quantum mechanically unstable, and will 
therefore decay with finite probability.

Here we have focused on the evolution of closed, spherically symmetric 
domain walls, which in general have a Schwarzschild mass and an
associated singularity.  In our analysis of the classical dynamics 
of these spacetimes, we have shown that four general classes of
behavior are possible:  a wall may be born with zero size, expand 
to a finite maximum radius and recollapse; it may be born with zero 
size and expand indefinitely; it may collapse from infinite size 
to a minimum radius and then reexpand; or it may collapse from 
infinite size to zero size.  In the case where the domain wall
expands indefinitely, it passes inside the black hole horizon,
avoids the singularity and creates a new, inflating universe which
is causally disconnected from the original spacetime, as first
pointed out by Blau et al. \cite{BGG}.  We have shown that this
process will occur naturally if the spherical domain wall is formed
with radius larger than the Hubble radius at formation.  

Our toy model to study the decay of such a domain wall-dominated
universe was chosen to be the limiting case where the Schwarzschild
mass of the wall vanishes.  In this case, we have seen that there
is an instanton which mediates the decay process, and that the
decay probability per unit time is
\begin{equation}
	P\sim\exp\left[-{2\over{\mu^2 G^3}}\right].
\end{equation}
Although the process is heavily suppressed, we emphasize that in
our scenario there are no competing processes; the domain wall
simply expands until it decays, whenever that may be.  The energy
of the wall will subsequently become thermalized and lead to a
hot big bang.

Our instanton calculation also predicts the classical state immediately
after the decay, namely a closed universe of smaller volume, consisting
of two regions of flat space meeting at a wall of energy.  The wall of
energy is the annihilation product of the two domain walls;  its geometry
is that of a 2-sphere with a radius $\sqrt2/\mu G$, or a factor
$1/\sqrt2$ smaller than the radius of the domain wall before the decay.

Discussion of the subsequent fate of this new universe is beyond the
scope of this paper.  But this closed universe now contains a form
of energy which obeys all the usual energy conditions, and therefore
recollapse may be expected.  If the wall of energy is in
a metastable state (``slow rolldown") then recollapse may be
delayed greatly, however, and this universe might expand to
a much greater radius before recollapse.

Our instanton calculation has also raised some interesting questions
about the overall nature and validity of the instanton approach to
quantum gravity.  In general, one would like to find a Euclidean
instanton which, in addition to being the least-action solution 
of the Euclidean field equations, satisfies two conditions:  
(i) it contains slices which are isomorphic to static slices in 
both the final and the initial state of the tunneling process, 
and (ii) the hypersurface which interpolates between these two 
slices has a unique trajectory.  However, as in our calculation, 
one finds that this is not always possible; in such cases, it is 
not clear how to proceed.  We have followed the rule of
Farhi, Guth and Guven \cite{FGG}, to evolve the interpolating
hypersurface on a 2-sheeted ``pseudomanifold" made of two instantons
glued together over a common region,
wherein the first sheet contains the final slice, and the second sheet
which contains the initial slice.  The two parts of the evolution 
are joined on a surface in the common region.  An 
alternate approach, which has become standard practice (see, for
example, \cite{MJP1,andrew}), is simply to subtract the action of
the instanton which mediates ``tunneling from nothing" of the initial 
state, from that of the instanton which describes ``tunneling from 
nothing" of the final state, and take this result as the action of
the interpolation.  This approach does not even require that the
two instantons have a single hypersurface which can be identified.
The two approaches give identical answers in the present
case, although the reasons for this are unclear.

Furthermore, our instanton is a saddlepoint, rather than a minimum,
of the Euclidean action for the class of field configurations we
have considered, as invariably happens in quantum gravity.  Since
there do not appear to be any extrema of lesser action, we have
taken our solution to be the dominant contribution to the decay
process.  However, we have not proven that this is the case. 

In order to avoid the ambiguities in the instanton calculation,
we turn directly to the canonical quantization of the
the minisuperspace model corresponding to the domain wall spacetime
under consideration.  Since the physical interpretation of the decay
process is that of a second domain wall being nucleated in one half 
of the existing spacetime, and the two walls tunneling towards each
other, there will be two degrees of freedom in the model.  We
will carry out the canonical quantization of such a two-wall spacetime
in the next paper \cite{paperII}, and we will find a significantly
different result for the quantum decay.

\acknowledgements

This research was supported in part by the National Science Foundation
under Grant Nos.~PHY94-07194 and PHY90-08502\@.  We are grateful to
Andrew Chamblin for a number of helpful conversations.

%%FIGURE CAPTIONS

\end{document}